\magnification=1200\hsize=14truecm\vsize=21truecm\parindent=0mm
\def\*{\vskip3mm}
{\bf Comments on }
\*
{\it On the application of the Gallavotti--Cohen fluctuation relation
to thermostatted steady states near equilibrium}
\*
by Evans, Searles, Rondoni in cond-mat/0312353 (third resubmission)
\*
\centerline{\it Author of the comments: Giovanni Gallavotti}
\centerline{I.N.F.N., Fisica, Roma 1}
\*\*
{\bf Abstract:} {\it The various versions of cond-mat/0312353
criticize results obtained by the author and coworkers in the last
decade. I have received requests to comment on the paper and the
comments are collected here, including some that I tried to point in
the course of discussions that can be found in the web page
http://ipparco.roma1.infn.it}
\*\*

The paper shows that a result by other Authors, [4], holds only at
large forcing while it does not hold at equilibrium or near it. A very
interesting and extremely surprising statement, if true.  I think that
the paper fails to achieve its purposes and I write a few comments.
\*

(1) There seems to be something inconsistent (possibly a matter of
notations) in the notions of isoenergetic and volume preserving
systems. Phase space can instantaneously contract (and expand) {\it
even} in isoenergetic systems, see p. 5 bottom, if isoenergetic means
at constant energy. And on p. 7 the {\bf A.} seem to contradict the
statement on p. 5 by saying that for constant energy dynamics the
extension leads to contradictions (which I understand means that phase
space can contract even in constant energy dynamics). This could be
made consistent, I believe.
\*
(2) On p. 6 the {\bf A.} attribute to ref. [4] the relation in eq. (2)
(see p. 6 top): however [4] does not even mention the relation in
eq. (2).
\*
(3) On p. 9 the {\bf A.} claim again that the ref. [4] result in
eq. (6) is equivalent to the relation in eq. (2): the latter however
is obviously false if $<\sigma>=0$: a case in which the two relations
cannot be equivalent being different. 
\*
(4) On p. 11 the {\bf A.} say that the checks on Chaotic Hypothesis are
''{\it somewhat circular}'': however the tests are performed on
systems which, surely, are not Anosov systems. The Chaotic Hypothesis
states that although certain systems are not Anosov they should share
most of their properties; hence the tests are meaningful (and
necessary). Therefore I cannot understand the {\bf A.'}'s statement.
\*
(5) On p. 11 the {\bf A.} seem surprised ({\it i.e.} find {\it
 problematic}) that the Fluctuation Relation is easier to detect at
 large forcing.  This is so, I believe, because at small forcing the
 slope is proportional to the dissipation $<\sigma>$ and therefore it
 is almost $0$: which the {\bf A.} might not see because they claim
 that the Fluctuation Relation is given by formula (2) which gives a
 {\it dimensional} slope equal to $1$: however in physical units that
 $1$ is {\it very} small at small forcing so that it is difficult to
 measure. In my view an important contribution of [4] was precisely to
 have identified the correct unit for measuring the dissipation:
 possibly a failure to realize this may have provided motivation for
 several papers that circulate.
\*
(6) On p. 11 bottom the {\bf A.} ''{\it have the impression that
distance from equilibrium does not play an essential role...}'':
however the result is completely different at zero dissipation if one
insists in writing it, in a not dimensionless form, as in formula (2)
(where an important physical quantity is ``arbitrarily'' set to $1$).
\*
(8) on p. 11 the {\bf A.} inquire about the domain of applicability of
eq. (2) and of the Chaotic Hypothesis: again I am afraid that this is
not properly stated nor done because the {\bf A.} require to check
formula (2) in cases with $<\sigma>$ close to $0$ while the formula is
{\it wrong} if $<\sigma>$ is exactly $0$: in other words the units
used in the measurement do not seem appropriate. I am puzzled because
this reminds me of wondering why a scale does not tilt if we add a
microscopic grain on one of the plates.
\*
(9) I skip comments on sect. 3 because some of its contents seem to
have been already commented in the literature to which the paper
refers (the second of ref. [36]) as trivial. It is possible that it
contains new results but a discussion of the (second) reference [36]
seems necessary to put this section into context.
\*
(10) On p. 18 the {\bf A.} say again that the fluctuation relation in
ref. [4] is incorrect and that it should be so because it does not
match eq (2). But eq (2) is wrong (in the form written, with no
conditions on $A$) and the {\bf A.} seem to be almost the only ones to
write it: certainly it is not in the papers that the {\bf A.}
criticize (ref [4], [37]).  The syllogism that follows is incorrect
because the premises are wrong.  The wrong assumptions are that
eq. (2) holds and that ref. [4] claims so: false (or else the {\bf A.}
should indicate where such an absurd claim is made).
\*
(11) On p. 19 the {\bf A.} seem to criticize their own equation
claiming that its derivation must have assumed some invalid
property. However the eq.  (2) has neither been claimed (in the form
written) nor proved by reference [1] nor by the authors of the other
references [4],[37],[28],[26] there or elsewhere. And since it is
obviously false, as the {\bf A.}  recognize, I cannot understand the
the purpose and the meaning of the comment.
\*
(12) On p. 20 The {\bf A.} claim (bottom of page) that ''{\it
Nevertheless the division by $e_f$ does not seem to be an essential
ingredient of the proof in [26].... Assuming that this is the
case....}'': however in a proof one cannot {\it assume the
conclusion}. It is obvious that $e_f\ne0$ is {\it an essential
ingredient}; and not taking that into account leads them to the wrong
conclusion that equ. (2) holds in equilibrium cases. The {\bf A.}
quote and claim to follow Ruelle's derivation of the Fluctuation
Theorem: however they do not. The quoted derivation is correct and it
cannot lead to the wrong result that the {\bf A.} claim (see line 12
p. 21). Furthermore the purpose of this ``proof'' seems to be, if I
understand, to criticize the Fluctuation Theorem in [4]: since this is
a fine mathematical point the {\bf A.} should have referred to the
original mathematical proof: {\it but they do not even quote it}; on
p. 9 they copy lines from p. 963 of [4] stopping short of the last
line of the same page where the mathematical proof, earlier than the
paper they quote, is referred.
\*
(13) On p. 22 the {\bf A.} seem to have doubts on the boundedness of
phase space contraction in Anosov cases and they seem to think that
this is an assumption: that is not the case because what they call
``assumption'' in [26] is not such, but it is one of the main results
of Anosov. The involved discussion on p. 22 seems to show that the
{\bf A.} seem to have realized that something was wrong in their
application of Ruelle's method.  But they do not seem to have realized
(if I understand the logic there) that they {\it should prove} that
$p^*$ is infinite and, furthermore, give arguments to show that what
matters, {\it i.e.} the product $p^*<\sigma>$, {\it is positive}
although $<\sigma>=0$. There might be some confusion between unbounded
$\Phi$ and unbounded $p^*$: it can be that even if $\Phi$ is unbounded
in the full phase space of ``all possible motions'' yet $p^*$ is still
bounded! I cannot see this point discussed by the {\bf A.}. This also
depends strongly on what is meant by ``isokinetic''. Is that Gaussian
isokinetic? or other? In no dissipative case that I know of is the
computation of $p^*$ easy. Except in the isokinetic equilibrium case
where $p^*$ vanishes (contrary to what I understand from the {\bf
A.}'s).
\*
(14) The problem should, in fact, be formulated in more detail, at
least as far as I can see: which is the phase space on which $p$ would
be unbounded? They seem to have numerical evidence that $p^*$ is
infinite: I am not convinced, because in a numerical experiment all
quantities are certainly bounded so $p^*$ can only be seen to be
large: and what matters is not $p^*$ but $p^*$ {\it times} $<\sigma>$
which will be very close to $0$ (because in the infinite precision
limit it will be $0$ at least when the system approaches equilibrium):
hence the delicate question arises on how big is the product. I cannot
find it discussed in the paper.
\*
(15) In fact in the case of thermostatted systems the {\bf A.} say
that $p^*$ might be infinite. I take ``thermostatted'' to mean
constant kinetic energy: this is however difficult for me because the
{\bf A.} talk all the time of thermostatted systems as systems at
constant temperature; {\it but} they never seem to define temperature,
which is one of the main problems in nonequilibrium. If so the {\bf
A.} might have failed to take into account that in order that $p$
(forgetting about $p^*$) be unbounded the initial data must start in a
configuration in which two Lennard-Jones (or W-Ch) particles are {\it
exactly} at the same point. Such an experiment cannot be made (the
computer would declare an error and refuse to continue). The matter is
delicate and the {\bf A.} do not discuss it in detail.  In other words
it is true that $p$ is unbounded in phase space in some models: {\it
but} a datum with finite energy at time zero keeps finite energy if
the total kinetic energy is bounded (and this happens in spite of the
presence of forces working on the system) and therefore its trajectory
will never visit the places where $p$ is too large.  My confusion
becomes really deep when a few lines later the {\bf A.}  consider the
Nos\'e-Hoover thermostat and seem to claim that {\it irrespective of
the boundedness of the potential} $p$ is unbounded: in this case I
would think that with probability $1$ on the auxiliary variable
$\zeta$ and for all initial conditions the value of $p$ stays bounded
(by a value depending on the initial data of the particles) and I
would wish a discussion on why the {\bf A.} think differently (as they
seem to do).
\*
(16) Claiming that $p^*$ is infinite simply because in the {\it full
phase} space it is unbounded might remind non attentive readers of
claiming that in statistical mechanics the temperature is infinite
because there is a slight probability that the total kinetic energy is
above an arbitrarily prefixed value.
\*
(17) The same comment is likely to apply to the Nos\'e--Hoover
thermostat with unbounded potentials (apparently not considered by the
{\bf A.}): while $p$ is in $(-\infty,+\infty)$ in the space of all
possible motions, why is $p^* = \infty$ if one starts with a finite
energy configuration? again I would think that with probability $1$ on
$\zeta$ both $p$ and $p^*$ stay finite. This is apparently realized,
to some extent, by the {\bf A.} on p. 23 third paragraph: but they
seem to simply skip considering the problem that they raise and
proceed to discuss ''{\it other scenarios}''. A discussion would be
necessary, in my view, as this is an interesting point (although I do
not see it to be really relevant for criticizing the Chaotic
Hypothesis and the Fluctuation Theorem of ref. [4]).
\*
(18) On p. 23 the {\bf A.} conclude that the eq. (2), which is never
claimed in [4], does not apply to thermostatted equilibrium systems:
being wrong for symmetry reasons of course it does not apply, hence
this comment does not apply to the references to which the {\bf A.}
seem to refer. To which references does the comment apply?  or which
is its purpose?
\*
(19) On p. 26 the {\bf A.} deal with a proof that ref. [37] is
     possibly wrong. They begin by claiming that ref. [37] refers to
     ergostatted systems, which is not true as the word ergostatted
     does not even appear in the quoted paper, nor anything equivalent
     to it.  The paper [37] refers to Anosov systems. The {\bf A.}
     continue by claiming that eq. (2) makes incorrect predictions:
     however as stated above neither [4] nor [37] claim eq. (2): the
     latter equation appears {\it instead} claimed by the {\bf A.}'s
     at p. 4 line 4. The Chaotic Hypothesis cannot lead to eq. (2),
     obviously, at least not in the form quoted without conditions of
     validity on $A$.  The analysis continues on the basis of the
     assumption that eq. (2) holds (see p. 29 line 7). The conclusion
     of a lengthy and in my opinion obscure argument is on p. 31 where
     the {\bf A.} say that in accord with sec. 4 (which contained the
     incorrect proof of eq. (2)) fluctuations in the Nose'-Hoover near
     equilibrium thermostatted dynamics are not consistent with
     eq. (2) at finite times. And therefore these systems must violate
     Chaotic Hypothesis. It looks like that the {\bf A.} have perhaps
     proved that eq. (2) is untenable. Which is a priori obvious and
     it did not require 31 pages to achieve: it remains mysterious to
     me why the {\bf A.} think that what they have discussed has
     anything to do with [37].
\*
(20) The first conclusion On p. 33 once more deals with eq. (2) and
     has therefore little to do with the Chaotic Hypothesis: because
     eq. (2) is proposed by the {\bf A.} and it does not follow from
     the Chaotic Hypothesis nor from [4].
\*
(21) Then the {\bf A.} claim that they cannot verify on computers
     eq. (2): once more this seems to me to be self criticism as that
     equation is proposed by them, see p. 4 line 4. This is very
     confusing and one is left to wonder what the {\bf A.} really
     think about thermostatted or barostatted (an undefined word as
     far as I could see). They continue by repeating that verification
     of eq. (2) becomes more and more difficult the closer one is to
     equilibrium. This might be due to a basic misunderstanding: that
     $A$ is a dimensional quantity whose natural unit of measure is
     $<\sigma>$ (and this is one of the key points of [4]) and
     therefore it becomes very difficult to observe when the latter
     quantity approaches $0$.  Why the reason is not the same as why a
     scale does not tilt if one adds a microscopic grain on one plate?
\*
(22) The {\bf A.} should explain what is wrong in the paper [28] which
     contradicts their (apparently main) conclusion at the bottom of
     p. 33.  In fact the paper deals with a system which is
     mathematically Anosov, and mathematically thermostatted. If the
     {\bf A.} read again the paper [28] that they quote they would
     probably recognize that their conclusion is incorrect at least in
     this case: here one studies a system as close as wished to
     equilibrium and one proves validity of the Fluctuation Theorem
     (the one corresponding to [4] with $<\sigma>>0$). And also
     validity of the Green-Kubo formula follows (they do not seem to
     quote the paper in which the author of [28] and his collaborators
     prove it). This shows that the computer experiments are very
     delicate near equilibrium and theory could be a much better
     guide.
\*
(23) The paper [28] (correct, I think, and the {\bf A.} do not claim
the contrary) provides a dramatic example of why the conclusion at
p. 33 ''{\it We interpret ...}'' can be grossly wrong.  Systems at
equilibrium can be Anosov and actually some of them ({\it e.g.} the
ones considered in [28]) provide the only known examples on how to
study in all detail several key matters debated by the {\bf A.}  ({\it
e.g.} the geodesic flow on a surface of constant negative curvature).
In such cases the proof in ref. [28] could be perfectly adapted even
to the equilibrium case and it would give a (trivial) symmetric
result! while the {\bf A.}'s formula (2), which they often seem to
attribute to [4], would give a result which is obviously wrong for
symmetry reasons.
\*
(24) In footnote [41] the {\bf A.} say that in [37] I do not assume
restrictions on ``thermostatting'' (correct) but it is also correct
that I do not use proportionality of the instantaneous flux and phase
space contraction. I am afraid that the {\bf A.}  criticize a
statement that I did not make or a property that I did not assume.
Here there is a dep problem: how does one define the duality between
fluxes and forces? this is usually defined phenomenologically. However
in the general frame of the Chaotic Hypothesis one can try a general
definition and that is what is attempted in [37] and in the subsequent
paper by Ruelle and myself (that the {\bf A.} do not quote).
\*
(25) I am very confused by this paper: however in my opinion, because
of the above comments, it certainly achieves one desirable goal: to
make clear the difference between the results in ref. [4] and the ones
the {\bf A.}  describe in sect. 3 and later: however this had been
already discussed in the second reference [36].
\*

{\bf References:} numbered as in cond-mat/0312353. Should the version
of cond-mat/0312353 be further amended the above comments refer to
version called {\bf v3}, now (26 February 2004) current and
downloadable from the abstract page.

%Giovanni Gallavotti, Roma 26 February 2004

\end